\documentclass[a4paper, 11pt]{article}
\usepackage[margin=2.5cm]{geometry}
\usepackage{latexsym}
\usepackage{graphicx}
\usepackage{epsfig}

 \renewcommand{\theequation}{\thesection.\arabic{equation}}
 \def\appendixa{
 \vskip 1cm
 {\bf \large Appendix: Geometric conventions and other formulae}
 \vskip 1cm
 \par
 \setcounter{equation}{0}
 \def\theequation{A.\arabic{equation}}
 }
 
\begin{document}

 \title{Weyl-Gauge Symmetry of Graphene}
 \author{Alfredo Iorio\thanks{E-mail: Alfredo.Iorio@mff.cuni.cz} \\
{\it Faculty of Mathematics and Physics, Charles University in Prague} \\
 {\it V Hole\v{s}ovi\v{c}k\'ach 2, 18000 Prague 8 - Czech Republic}}
   \date{\today}

\maketitle
\begin{abstract}
\noindent The conformal invariance of the low energy limit theory governing the electronic properties of graphene is explored. In particular, it is noted that the massless Dirac theory in point enjoys local Weyl symmetry, a very large symmetry. Exploiting this symmetry in the two spatial dimensions and in the associated three dimensional spacetime, we find the geometric constraints that correspond to specific shapes of the graphene sheet for which the electronic density of states is the same as that for planar graphene, provided the measurements are made in accordance to the inner reference frame of the electronic system. These results rely on the (surprising) general relativistic-like behavior of the graphene system arising from the combination of its well known special relativistic-like behavior with the less explored Weyl symmetry. Mathematical structures, such as the Virasoro algebra and the Liouville equation, naturally arise in this three-dimensional context and can be related to specific profiles of the graphene sheet. Speculations on possible applications of three-dimensional gravity are also proposed.
\end{abstract}

\bigskip



\bigskip

\section{Introduction}
\setcounter{equation}{0}

In the low energy limit the electronic properties of graphene are well described by massless Dirac spinors $\Psi$ in two space and one time dimensions, hence an analog of a relativistic system but with characteristic velocity given by the Fermi velocity $v_F$ rather than the speed of light $c$ \cite{review2009}
\begin{equation}\label{first}
    A = i \hbar v_F \int d^3 x \bar{\Psi} \gamma^a \partial_a \Psi \;,
\end{equation}
where $a = 1,2,3$ is a (flat) spacetime index. This action is scale and conformally invariant\footnote{In the next Section we shall explain in detail the related concepts of scale, conformal and Weyl symmetries, both rigid and local, especially for the Dirac action. For the general case see \cite{Iorio:1996ad}.}.

The main messages of this paper are two, one general and one practical. The general suggestion is that {\it intrinsic} curvature of the two-dimensional graphene sheet within this special relativistic-like behavior naturally leads to a general relativistic-like description\footnote{The ``like'' suffix here and in the following refers to the fact that this is not a fundamental theory but rather an effective one, one consequence being that the limiting speed is $v_F$ rather than $c$. Nonetheless, besides that, considerations are along the same lines as for the fundamental theories.} of $\Psi$ regarded as a Dirac field in a {\it curved three-dimensional spacetime} back-ground
\begin{equation}\label{second}
    {\cal A} = i \hbar v_F \int d^3 x \sqrt{g} \bar{\Psi} \gamma^a \nabla_a \Psi \;,
\end{equation}
where $\nabla_a$ is the general covariant derivative, an important role being played by the {\it local} Weyl symmetry enjoyed by the electronic system. This suggestion, we shall see, comes very naturally when time is included in the analysis. The second, practical, main message of this paper is that  the local Weyl symmetry of the graphene action can simplify the analysis of its electronic properties to the extent of allowing to find specific shapes of the graphene sheet for which {\it the electronic local density of states (LDOS) is  the same as that for a planar sheet}, provided times and lengths are measured in accordance to the inner reference frame of the electronic system.

We find that when the bending of the graphene sheet gives rise to a constant intrinsic two-dimensional curvature, the three-dimensional metric (the one that includes time) is truly curved also in the time direction, although, for the cases specifically considered here, only in a very special way (conformally flat). Once such a general relativistic-like spacetime is accepted as physically meaningful, a special role is played by these conformally flat spacetimes
\begin{equation}\label{third}
    g^{(3)}_{\mu \nu} = \varphi (x) \eta^{(3)}_{\mu \nu} \;,
\end{equation}
because in that case the curved space action (\ref{second}) is {\it equal} to the flat space action (\ref{first}), provided the fields are related by a Weyl transformation
\begin{equation}
{\cal A} (\Psi ') = A(\Psi) \;,
\end{equation}
hence the Euler-Lagrange (Dirac) equations for $\Psi '$ in the curved case are equal to those for $\Psi$ in the flat case. As we shall see, this is so because of the special status of the massless Dirac action in any dimension. The three dimensions play a crucial role for other important matters, like for instance the particular form for the constraints for conformal flatness (it is the Cotton tensor rather than the Weyl tensor that needs to vanish) but this particular instance, ${\cal A} (\Psi ') = A(\Psi)$, is independent from the dimensions. This fact is known in field theory on curved spaces as an instance of {\it conformal triviality} \cite{birrellanddavies}. We shall revisit that instance and shall show that it has the consequence of an \textit{invariant electronic LDOS}. Our results do not seem to contradict existing studies of the nontrivial effects of curvature on the LDOS, such as, e.g, those reported in \cite{vozmediano}, for two reasons: first, the {\it ad hoc} choice of the shape made there can well be outside the class of Weyl invariant choices (surely it is not within the class identified here); second, the measuring procedure adopted there seems to be a non-invariant one (we shall be more explicit later on this point).

The paper is organized as follows. In Section 2 we shall recall the concepts of scale, conformal and Weyl symmetries in general and for the case of a massless Dirac field, and discuss the special case of conformal triviality and its impact on the Green functions. In Section 3 we shall relate the above to the case in point of the electronic properties of a graphene sheet and shall discuss the impact on the LDOS. In Section 4 we shall see which two-dimensional bending of the graphene sheet produces a case of conformal triviality in the three-dimensional spacetime and learn that these conditions are quite natural, i.e. constant intrinsic curvature of the two-dimensional sheet. This means three possibilities: i) the sheet is a plane; ii) the sheet is flat but not planar (we find that this corresponds to two-dimensional conformal factors that are harmonic functions); iii) the bent sheet is truly curved with a constant intrinsic curvature (we find that this corresponds to two-dimensional conformal factors that are solutions of the Liouville equation). There we notice that, although the system is truly a three-dimensional one, due to the nature of the metric used, two-dimensional conformal field theory structures, such as the Virasoro algebra or the Liouville equation naturally appear. Section 5 is devoted to collect our results and to discuss various possible directions for future investigations, in particular whether it makes sense for the physics of graphene to use the rich landscape of solutions of three-dimensional gravity, including black-holes.

\section{Scale, Conformal and Weyl Symmetries of the Dirac Action}
\setcounter{equation}{0}

Scale, conformal, and Weyl symmetries are intimately related but different. For the general case of theories in any dimension and with arbitrary spin the issue was discussed at length in \cite{Iorio:1996ad}. In this paper we want to focus on the special case of the massless Dirac field in two space and one time dimensions because it is relevant for graphene. To fully appreciate the peculiarities of this case, though, and for the sake of introducing notation and concepts let us start this Section by reviewing the results of \cite{Iorio:1996ad} and by putting them in the right perspective for the applications we intend to pursue.

Consider an action in $n$ dimensions for the field $\Phi_i$, where $i$ stands for a generic spin index, invariant under general coordinate transformations $x_\mu \to \tilde{x}_\mu (x)$ (diffeomorphisms) and local Lorentz transformations $x_a \to {\omega_a}^b x_b$
\begin{equation}\label{diffeogeneralaction}
    {\cal A} (g_{\mu \nu}, \Phi_i , \nabla_a \Phi_i) = \int d^n x \sqrt{g} {\cal L} (\Phi_i , \nabla_a \Phi_i) \;.
\end{equation}
This action could be in flat space in curvilinear coordinates or in a truly curved geometry. We implicitly introduced the Vielbein $e^a_\mu$ and its inverse $E_a^\mu$
\begin{equation}\label{vielbein}
    \eta_{a b} e^a_\mu e^b_\nu = g_{\mu \nu} \;\; \;\;  e^a_\mu E_a^\nu = \delta_\mu^\nu \;\; \;\; e^a_\mu E_b^\mu = \delta_b^a \;,
\end{equation}
the indices $\mu , \nu = 0, 1, ..., n-1$ that respond to diffeomorphisms (Einstein indices), while $a, b = 0, 1, ..., n-1$ respond to flat space transformations (Lorentz indices), $\eta_{a b} = {\rm diag}(+1, -1, ...)$, $\sqrt{g} \equiv \sqrt{|\det g_{\mu \nu}|}$ and the diffeomorphic covariant derivative that is
\begin{equation}\label{grcovdev}
    \nabla_\mu \Phi_i = \partial_\mu \Phi_i + {\Omega_\mu}_i^{\; j} \Phi_j  \;,
\end{equation}
with $\nabla_a = E_a^\mu \nabla_\mu$, ${\Omega_\mu}_i^{\; j} =  \frac{1}{2} \omega_\mu^{a b} (J_{a b})_i^{\; j}$, $(J_{a b})_i^{\; j}$ the appropriate form of the generators of the Lorentz transformations\footnote{E.g., if $\Phi_i = \phi$ is a scalar field then $J_{a b} = 0$, while if $\Phi_i = \Psi_\alpha$ is a Dirac spinor, the case we shall be dealing with, then $(J_{a b})_\alpha^{\; \beta}
= \frac{1}{4} [ \gamma_a , \gamma_b ]_\alpha^{\; \beta}$, etc..} and
\begin{equation}\label{spinconngeneral}
    {\omega_\mu}^a_{\; b} = e^a_\lambda (\delta^\lambda_\nu \partial_\mu + \Gamma_{\mu \nu}^\lambda) E^\nu_b \;,
\end{equation}
is the spin connection obtained by requiring the full covariant derivative of the Vielbein to be zero (metricity condition)
\begin{equation}\label{fullcovdevviel}
    \nabla_\mu e^a_\nu = \partial_\mu e^a_\nu - \Gamma_{\mu \nu}^\lambda e^a_\lambda +  \omega_{\mu \; b}^a e^b_\nu = 0 \;,
\end{equation}
where $\Gamma_{\mu \nu}^\lambda$ is the Christoffel connection. Torsion is defined as $T^\lambda_{\;\; \mu \nu} = \Gamma^\lambda_{\;\; [ \mu  \nu]} = E_{\;\; a}^\lambda T^a_{\;\; \mu \nu}$ where the expression in terms of Vielbein is $T^a_{\;\; \mu \nu} = ( \partial_{[\mu} \; e_{\nu]} + \omega_{[\mu} \; e_{\nu]} )^a $, see the Appendix. At this point it is not required that $T^\lambda_{\;\; \mu \nu} = 0$ and to keep track of that might become important if topological defects described by a Cartan-Einstein model of gravity with torsion (dislocations) need to be included \cite{KatVol92}, \cite{Kleinert}.

Due to diffeomorphism invariance of (\ref{diffeogeneralaction}) a rigid scaling of the coordinates can be transferred to the Vielbein\footnote{Under diffeomorphisms
\begin{equation}\label{diffeoviel}
    \tilde{e}^a_\mu (\tilde{x})
    = \frac{\partial x^\lambda}{\partial {\tilde x}^\mu} e^a_{\lambda} (x) \;,
\end{equation}
which, requiring $\tilde{x}_\mu \equiv e^{- \sigma} x_\mu$, gives $\tilde{e}^a_\mu (\tilde{x}) = e^\sigma e^a_\mu (x)$, that is the same result obtained with $\tilde{x}_\mu = x_\mu$ and requiring $\tilde{e}^a_\mu (\tilde{x}) \equiv e^\sigma e^a_\mu (x)$. The second choice, though, hides an important difference, namely that $\tilde{x}$ in $\tilde{e}^a_\mu$  is indeed just $x$, hence the transformation now does not involve the coordinates but only the fields  $e^a_\mu$.} (or, equivalently, to the metric). It is then clear then that when the flat space limit of ${\cal A} (g)$ in (\ref{diffeogeneralaction}), say
\begin{equation}\label{flatgeneralaction}
    {\cal A} (\eta_{a b}, \Phi_i , \partial_a \Phi_i) = \int d^n x {\cal L} (\Phi_i , \partial_a \Phi_i) \;,
\end{equation}
is symmetric under rigid scaling
\begin{equation}\label{scalegeneral}
    x^a \to e^\sigma x^a \;\; {\rm and} \;\; \Phi_i \to e^{d_\Phi \sigma} \Phi_i \;,
\end{equation}
where $d_\Phi$ is the scale dimension of the field dictated by the kinetic term, then ${\cal A} (g)$ is symmetric under
\begin{equation}\label{weylgeneral}
    e^a_\mu \to e^\sigma e^a_\mu \;\; {\rm and} \;\; \Phi_i \to e^{d_\Phi \sigma} \Phi_i \;.
\end{equation}
The transformations (\ref{weylgeneral}) are called {\it rigid} (being $\sigma$ constant) Weyl transformations. They are called {\it local} Weyl transformations if $\sigma(x)$.

Let us now recall that often Poincar\`e and scale invariant actions are also invariant under the full conformal group of transformations, namely, on top of the Poincar\`e and scale transformations, they are invariant under those transformations for which the scaling parameter is\footnote{Other choices often used are $\sigma_{\rm SC} = - \frac{1}{2} \ln (1 + 2 c x + c^2 x^2)$ and $\sigma_{\rm SC} = - \ln (1 + 2 c x + c^2 x^2)$ (and sometimes $c_a \to - c_a$, as in \cite{Iorio:1996ad}). They are all equivalent to the application to $x_a$ of an inversion followed by a translation by $c_a$, followed by an inversion.}
\begin{equation}\label{specialconf}
    \sigma_{\rm SC} (x) = \frac{1}{2} \ln (1 + 2 c x + c^2 x^2) \;,
\end{equation}
where $c_a$ is a constant vector parameter, and SC stands for \textit{special conformal} transformations. This is the case of the massless Dirac action of our concern here but it is not always true.

The exact criterion to establish when it is true that scale invariance of the flat space action implies full conformal invariance of the latter invokes the local Weyl invariance of the curvilinear action, as one might suspect by noticing that $\sigma_{\rm SC}$ is coordinate dependent. The general procedure is presented in detail in \cite{Iorio:1996ad} and we shall not do it again here but it is crucial for us to understand the special status of the Dirac action in any dimension with respect to fields of integer spin.

Thus, let us consider the case of a general $x$-dependent scaling parameter $\sigma(x)$, i.e. not constrained to be $\sigma_{\rm SC}$. For $\sigma(x)$ the Weyl transformations (\ref{weylgeneral}) are not symmetries of ${\cal A} (g)$ in (\ref{diffeogeneralaction}), just like the scale transformations (\ref{scalegeneral}) when $\sigma(x)$, in general, are no longer symmetries of ${\cal A}(\eta)$ in (\ref{flatgeneralaction}). To make ${\cal A}(g)$ invariant we need to gauge it {\it \`a la} Weyl (these considerations now presume torsionless connections)
\begin{equation}\label{weylgaugedA}
    {\cal A} (g_{\mu \nu}, \Phi_i , \nabla_a \Phi_i) \Rightarrow {\cal A}_W (g_{\mu \nu}, \Phi_i , {\cal D}_a \Phi_i) \;,
\end{equation}
where
\begin{equation}\label{virialcovdev}
{\cal D}_\mu \Phi_i = \nabla_\mu \Phi_i + (\Lambda_\mu^{\; \nu})_i^{\; j} W_\nu \Phi_j
\end{equation}
and
\begin{equation}\label{virial}
    (\Lambda_{\mu \nu})_i^{\; j} = d_\Phi g_{\mu \nu} \delta_i^{\; j} + (J_{\mu \nu})_i^{\; j} \;,
\end{equation}
with
\begin{equation}\label{weylfieldandcovdev}
    W_\mu \to W_\mu - \sigma_\mu \; \; {\rm and} \; \; {\cal D}_\mu \Phi_i \to e^{d_\Phi \sigma} {\cal D}_\mu \Phi_i \;,
\end{equation}
under (\ref{weylgeneral}) with $\sigma(x)$. Here $\sigma_\mu \equiv \partial_\mu \sigma$ and $J_{\mu \nu} = e^a_\mu J_{a b} e^b_\nu$. As for usual gauge theories, the action is {\it modified} to a new one that is invariant: ${\cal A}_W (g) \to {\cal A}_W (g)$ under (\ref{weylgeneral}) with $\sigma(x)$. If and only if this produces an ${\cal A}_W (g)$ where $W_\mu$ only appears in a combination, say it $\Omega_{\mu \nu} [W_\lambda]$, such that in the flat space limit $\Omega_{a b} [ \sigma_c ] = 0 $ has $\sigma_{\rm SC}$ in (\ref{specialconf}) as the only solution, then the flat space action (\ref{flatgeneralaction}) is conformal invariant. In general this only happens for certain combinations of the spin of the fields and of the dimension of the spacetime.

What just said needs only a little adjustment in relation to the two-dimensional case, where the $\sigma_{\rm SC}$s are many more than just (\ref{specialconf}) as all harmonic $\sigma (x)$s (i.e. the infinite solutions of $\partial_a \partial^a \sigma = 0$) correspond to conformal transformations. The procedure in this case gives as necessary and sufficient condition for a scale symmetric ${\cal A} (\eta)$ to be fully conformal invariant that $W_\mu$ enters ${\cal A}_W (g)$ as $\Omega^\mu_\mu$. Special attention is needed for the scalar fields, as pointed out in \cite{jackiw2dweyl}.

In this respect, the status of the massless Dirac action\footnote{Although for slightly different reasons, what follows is also true for scale invariant actions of fields of any half-integer spin (Rarita-Schwinger).} in any dimension is very special because\footnote{For the moment, we shall take $\hbar = 1 = v$ with $v$ a parameter with the dimensions of velocity, for graphene it is $v = v_F$, the Fermi velocity, for truly relativistic Dirac it is, of course, $v = c$.}
\begin{equation}\label{diracgeneral}
     {\cal A}_W (g_{\mu \nu}, \Psi , {\cal D}_a \Psi) = i \int d^n x \sqrt{g} \; \bar{\Psi} \gamma^a E^\mu_a
     (\nabla_\mu + \Lambda_\mu^{\; \nu} W_\nu) \Psi = {\cal A} (g_{\mu \nu}, \Psi , \nabla_a \Psi) \;,
\end{equation}
due to
\begin{equation}\label{virialzero}
    \gamma^\mu \Lambda_{\mu \nu} = d_\Psi \gamma_\nu + \gamma^\mu J_{\mu \nu} = 0 \;,
\end{equation}
where we used $d_\Psi = (1 -n)/2$ and the definition of the Lorentz generators $J_{\mu \nu} = \frac{1}{4} [\gamma_\mu, \gamma_\nu]$. This means that not only the Dirac action in flat space
\begin{equation}\label{diracflat}
    {\cal A} (\eta_{a b}, \Psi , \partial_a \Psi) = i \int d^n x \bar{\Psi} \gamma^a \partial_a \Psi \;,
\end{equation}
is conformal invariant but that the curvilinear or truly curved space action
\begin{equation}\label{diraccurve}
 {\cal A} (g_{\mu \nu}, \Psi , \nabla_a \Psi)
= i \int d^n x \sqrt{g} \; \bar{\Psi} \gamma^a E^\mu_a (\partial_\mu + \frac{1}{2} \omega_\mu^{\; b c} J_{b c}) \Psi \;,
\end{equation}
is local Weyl invariant, as can be also seen directly by transforming (\ref{diraccurve}) under (\ref{weylgeneral}) with $\sigma (x)$ obtaining
\begin{equation}\label{directweyl}
    {\cal A} (g_{\mu \nu}, \Psi , \nabla_a \Psi) \to {\cal A} (g_{\mu \nu}, \Psi , \nabla_a \Psi) + i \int d^n x \sqrt{g} \; \bar{\Psi} \gamma^\mu \Lambda_\mu^{\; \nu} \sigma_\nu \Psi = {\cal A} (g_{\mu \nu}, \Psi , \nabla_a \Psi) \;.
\end{equation}
Here it was used $\omega_\mu^{\; a b} \to \omega_\mu^{\; a b} + (e^a_\mu e^b_\nu - e^b_\mu e^a_\nu)$. This and other relevant formulae are in the Appendix.

In any dimension this is a very large symmetry that ${\cal A} (g_{\mu \nu}, \Psi , \nabla_a \Psi)$ enjoys because there are no restrictions to $\sigma(x)$ of any sort. This happens for ${\cal A} (g)$ as it stands which means that the theory is, we may say, ``intrinsically gauged'': ${\cal A}_W (g) = {\cal A}(g)$. We shall exploit this symmetry in what follows for the graphene action where a crucial role is played by the dimensionality of the problem, which is two space and one time dimensions. Let us see here what this symmetry means in physical terms in any dimensions, including the case of graphene.

When two metrics are related as $g'_{\mu \nu} = e^{2 \sigma (x)} g_{\mu \nu}$ the classical physics for the field $\Psi' = e^{d_\Psi \sigma (x)} \Psi$ in $g'_{\mu \nu}$ is precisely the same as the classical physics for the field $\Psi$ in $g_{\mu \nu}$  because ${\cal A}(g_{\mu \nu}, \Psi , \nabla_a \Psi) = {\cal A}({g'}_{\mu \nu}, \Psi' , {\nabla'}_a \Psi')$. Furthermore, we can choose
\begin{equation}
g'_{\mu \nu} \equiv \eta_{\mu \nu} \;,
\end{equation}
and obtain
\begin{equation}\label{corerelation1}
i \int d^n x \sqrt{g} \bar{\Psi} \gamma^a E_a^\mu \nabla_\mu \Psi = i \int d^n x \; e^{- (n -1) \sigma} \; \bar{\Psi} \gamma^a
(\partial_a - \frac{n - 1}{2} \sigma_a) \Psi = i \int d^n x \bar{\Psi}' \gamma^a \partial_a \Psi' \;,
\end{equation}
when
\begin{equation}\label{corerelation2}
    g_{\mu \nu} = e^{-2 \sigma (x)} \eta_{\mu \nu} \;\;\;\; {\rm and} \;\;\;\; \Psi = e^{\frac{(n-1)}{2} \sigma(x)} \Psi' \;,
\end{equation}
where we used the simple algebra illustrated in the Appendix to write the second form of the action.

We are then dealing with a conformally invariant field in a conformally flat spacetime, a case sometimes referred to as {\it conformal triviality}\footnote{Notice, once more, that for fields of half-integer spin conformal triviality needs not a coupling of the fields to curvature tensors to have local Weyl invariance, while this is necessary for integer spin.} \cite{birrellanddavies}. This means that the effects of curvature are null on the classical physics of a massless Dirac field $\Psi$ as in (\ref{corerelation2}) if the spacetime is only curved in a conformally flat fashion: if the metrics can only be the conformally flat we know from here that the classical physics is once and for all governed by the flat space action. Of course this is not always true as there is no reason in general to consider only conformally flat metrics, but when this happens there is the ``magic'' consequence that although the spacetime is curved this has no effects whatsoever on the classical physics of the system, provided the field is properly redefined.

What for the quantum case? If we define the flat propagator as usual
\begin{equation}\label{flatpropagator}
    i S ' (x_1, x_2) \equiv \; '\langle 0| \Psi ' (x_1) \bar{\Psi}' (x_2) |0 \rangle ' \;,
\end{equation}
where
\begin{equation}\label{flatpropequation}
    i \not\!\partial_{x_1} S ' (x_1, x_2) = \delta^{n} (x_1, x_2) \;,
\end{equation}
we can see the effect of the Weyl transformations by considering the unitary operator that implements the transformations quantum mechanically, i.e.
\begin{equation}\label{bogoliubovfield}
    \Psi (x) = U \Psi' (x) U^{-1} = e^{\frac{n-1}{2} \sigma(x)} \Psi'(x) \;.
\end{equation}
By writing $U = e^B $ and by using $[\Psi'(x), {\Psi'}^\dagger (y)]_{+} = \delta^n (x-y)$, and from
\begin{eqnarray}\label{operatorB}
U \Psi' (x) U^{-1} & = & \Psi'(x) + [B,\Psi'(x)] + \frac{1}{2} [B,[B,\Psi'(x)]] + \cdots  \nonumber \\
& = & \Psi'(x) + \frac{n-1}{2} \; \sigma(x) \Psi'(x) + \frac{1}{2} \left( \frac{n-1}{2} \; \sigma(x) \right)^2 \Psi'(x) + \cdots \;,
\end{eqnarray}
we easily obtain
\begin{equation}\label{U}
    U = \exp \left\{ \frac{1 - n}{2} \int d^n y \sigma(y) {\Psi'}^\dagger (y) \Psi'(y) \right\} \;,
\end{equation}
i.e., $B^\dagger = - B$ and $U(- \sigma) = U^{-1} (\sigma)$. Notice also that $[B, \gamma^a] = 0$, hence $U \gamma^a U^{-1} = \gamma^a$ and
$\bar{\Psi} (x) = U \bar{\Psi}' (x) U^{-1} = e^{\frac{n-1}{2} \sigma(x)} \bar{\Psi}'(x)$. The vacuum associated to the Weyl transformed fields is of course
\begin{equation}\label{Uvacuum}
    |0\rangle = U |0\rangle' = \exp \left\{ \frac{1 - n}{2} \int d^n y \sigma(y) {\Psi'}^\dagger (y) \Psi'(y) \right\} |0\rangle' \;,
\end{equation}
that shows the typical condensate structure of a quantum field in a curved (classical) background \cite{TFDBH}. Now, as discussed at length, e.g., in \cite{TFDBH}, we can do two things: either we consider
\begin{eqnarray}\label{directgreen}
    i S  (x_1, x_2) & \equiv & \langle 0| \Psi  (x_1) \bar{\Psi} (x_2) |0 \rangle \label{directgreen1} \\
    &=& \; '\langle 0| U^{-1} U \Psi ' (x_1) U^{-1} U \bar{\Psi}' (x_2) U^{-1} U |0 \rangle ' \label{directgreen2} \\
    &=&  \; '\langle 0| \Psi ' (x_1) \bar{\Psi}' (x_2) |0 \rangle '  = i S ' (x_1, x_2) \label{directgreen3} \;,
\end{eqnarray}
or we consider
\begin{eqnarray}
    i S^\sigma (x_1, x_2) \equiv \; '\langle 0| \Psi  (x_1) \bar{\Psi} (x_2) |0 \rangle ' & = & e^{\frac{n-1}{2} (\sigma(x_1) + \sigma(x_2))} \;
    '\langle 0| \Psi ' (x_1) \bar{\Psi}' (x_2) |0 \rangle '  \label{crossgreen1} \\
    i S^{- \sigma} (x_1, x_2) \equiv \; \langle 0| \Psi ' (x_1) \bar{\Psi}' (x_2) |0 \rangle & = & e^{\frac{1 - n}{2} (\sigma(x_1) + \sigma(x_2))} \; '\langle 0| \Psi ' (x_1) \bar{\Psi}' (x_2) |0 \rangle ' \;, \label{crossgreen2}
\end{eqnarray}
notice that $\langle 0| \Psi ' (x_1) \bar{\Psi}' (x_2) |0 \rangle \to \; '\langle 0| \Psi (x_1) \bar{\Psi} (x_2) |0 \rangle '$ when $\sigma \to - \sigma$, as it should be from $U(- \sigma) = U^{-1} (\sigma)$. Summarizing:
\begin{eqnarray}\label{green1}
    S(x_1, x_2)  & = & S'(x_1, x_2) \\
    S^\sigma (x_1, x_2) & = & e^{\frac{n-1}{2} (\sigma(x_1) + \sigma(x_2))} S'(x_1, x_2) \label{green2} \;,
\end{eqnarray}
with $S^\sigma (x_1,x_2)$ the Green function for the curved Dirac operator
\begin{equation}\label{propequation}
    i \not\!\nabla_{x_1} S^\sigma  (x_1, x_2) = \frac{1}{\sqrt{g}} \delta^{n} (x_1, x_2) \;,
\end{equation}
as proved in the Appendix.

The physical meaning of the two procedures is clear. In the first case, Eqs. (\ref{directgreen1})-(\ref{directgreen3}), the measurements are made within the frame of reference that is co-moving with the particles described by the field $\Psi$, hence the effects of curvature are completely removed also at the quantum level. In the second case, Eqs. (\ref{crossgreen1}) and (\ref{crossgreen2}), the measurements are made by an observer that sees the (quantum) effects of curvature in the form of a condensate in the vacuum, Eq. (\ref{Uvacuum}).

We faced this sort of situation in \cite{TFDBH} where we dealt with vacuum expectations of the kind $\langle 0(\epsilon)| N |0(\epsilon)\rangle$ (or viceversa $\langle 0| N (\epsilon) |0 \rangle$), with $U(\epsilon) N U^{-1}(\epsilon) = N(\epsilon)$ and $U(\epsilon)|0\rangle = |0(\epsilon)\rangle$, where $|0(\epsilon)\rangle$ is the vacuum for a scalar field, e.g., in a Schwarzschild spacetime, with $\epsilon$ a function related to the acceleration $a$ or to the surface gravity $\kappa$ (near the horizon the space-time is always Rindler, see, e.g., \cite{Castorina:2008gf}). The number operator $N$ is that of the same field in an inertial frame, i.e. a frame that is not freely falling into the black-hole with acceleration $a$. In other words, $N$ represents a particles' ``counter'' placed asymptotically far-away from the horizon in a region that can be approximated as flat. It is this counter that sees the condensate structure of $|0(\epsilon)\rangle$
\begin{equation}\label{Numezawa}
    \langle 0(\epsilon)| N |0(\epsilon)\rangle = \sinh^2 \epsilon (p) = \frac{1}{e^{\beta \Omega} - 1} \;,
\end{equation}
where $p = (\Omega, \vec{k})$ is the four momentum in the curved frame, $\beta = 1/T$ is the inverse temperature, the $\sinh^2 \epsilon (p)$ descends from the algebraic structure of $U(\epsilon)$ and the relation of the latter to the Bose distribution comes from the minimization of the free energy \cite{TFDBH}. If, on the other hand, we let our particle counter fall into the black hole, this will not see any horizon, nor there will be a way for it to detect the acceleration, as the principle of equivalence dictates. Mathematically this translates to
\begin{equation}\label{Numezawadirect}
    \langle 0(\epsilon)| N (\epsilon)|0(\epsilon)\rangle = 0 \;,
\end{equation}
that is precisely what we get in the absence of any gravitational/acceleration effect, $\langle 0 | N |0 \rangle = 0$.

Returning to our case, if we think of $\Psi \bar{\Psi}$ as our ``counter'', we can apply the same logic just illustrated, i.e. $S(x_1, x_2)   =  S'(x_1, x_2)$ means that we see no effects of curvature, while $S^\sigma (x_1, x_2)$ contains the information on the quantum vacuum condensate. Two instances are important here. First, the physical picture can only become fully clear when the particular (conformally flat) metric $g_{\mu \nu} = e^{-2 \sigma} \eta_{\mu \nu}$ has been specified. When we can do that we can give meaning to lengths and times measured according to the line element $ds^2 = e^{-2 \sigma} (dt^2 - d{\vec{x}}^2)$, hence we can appreciate the meaning of the invariance $S(x_1, x_2)   = S'(x_1, x_2)$. What this means practically (i.e. what needs to be done to the measuring apparatus) depends upon $\sigma$. Second, in \cite{TFDBH} we have been dealing with a truly field theoretical setting, i.e. with an infinite number of degrees of freedom and the related infinite volume limit. In that case the Bogoliubov operators $U(\epsilon)$ were not well defined in the limit and unitarity was lost (on this see, e.g., \cite{qVN}). For the applications to graphene we can safely take the view that unitarity of $U(\sigma)$ is never lost because the size of the sample is always finite and the number of degrees of freedom cannot be infinite because to excite more and more degrees of freedom we would need higher and higher energy than that necessary for the ``Dirac-like'' approximation to hold (see next Section).

A last remark on the quantum regime is related to the fact that what just said seems to apply to any $U$, i.e. not just to the $U(\sigma)$s in (\ref{U}) that generate the Weyl transformations hence are symmetries of the classical theory. To select which $U$s are to be used one can look at the path integral version of the Green functions (see, e.g., \cite{ramond}) and use the classical Weyl-symmetry of the action ${\cal A}(\Psi, \sigma) = {\cal A}(\Psi')$
\begin{eqnarray}
S^{PI} (x_1, x_2) & \equiv & \frac{\int {\cal D} \Psi {\cal D} \bar\Psi \; \Psi(x_1) \bar\Psi(x_2) \; \exp\{ i {\cal A}(\Psi, \sigma) \}}
{\int {\cal D} \Psi {\cal D} \bar\Psi \exp\{ i {\cal A}(\Psi, \sigma) \}}  \label{PIGreen1} \\
& = & e^{\frac{n-1}{2} (\sigma(x_1) + \sigma(x_2))} \frac{\int {\cal D} \Psi' {\cal D} \bar\Psi' \; \Psi'(x_1) \bar\Psi'(x_2) \; \exp\{ i {\cal A}(\Psi') \}}{ \int {\cal D} \Psi' {\cal D} \bar\Psi' \exp\{ i {\cal A}(\Psi') \}}  \label{PIGreen2} \\
& \equiv & e^{\frac{n-1}{2} (\sigma(x_1) + \sigma(x_2))} S' (x_1, x_2) \label{PIGreen3} \;.
\end{eqnarray}
Here, since $\sigma(x)$ is an external field (see later), we used ${\cal D} \Psi = e^{\frac{n-1}{2} \sigma(x)} {\cal D} \Psi'$, similarly for ${\cal D} \bar\Psi$, hence, in the ratio the overall factor $e^{\frac{n-1}{2} (\sigma(x) + \sigma(y))}$ cancels. From the above we see that the Green function computed via standard path integral methods, $S^{PI}$, is indeed $S^\sigma$ in (\ref{crossgreen1}). This is so because we can as well read (\ref{PIGreen2}) as
\begin{equation}\label{linkgreen}
    \frac{\int {\cal D} \Psi' {\cal D} \bar\Psi' \; \Psi(x_1) \bar\Psi(x_2) \; \exp\{ i {\cal A}(\Psi') \}}{\int {\cal D} \Psi' {\cal D} \bar\Psi' \exp\{ i {\cal A}(\Psi') \}} \equiv  \; '\langle 0| \Psi  (x_1) \bar{\Psi} (x_2) |0 \rangle ' \;.
\end{equation}
Nonetheless, after having exploited the classical symmetries in the path integral Green functions we can then use the operator approach to find the $U$s corresponding to these symmetries, for us here the $U(\sigma)$s in (\ref{U}), and as well construct the invariant Green function as in (\ref{directgreen2}).

Let us now compare Weyl gauge symmetry to standard gauge symmetry. To this end consider the Dirac action
\begin{equation}\label{diracgauge}
    {\cal A} (\eta_{a b}, \Psi , A_a ) = i \int d^n x \bar{\Psi} \gamma^a ( \partial_a + A_a) \Psi
\end{equation}
invariant under
\begin{equation}\label{standardgaugelocal}
    \Psi \to \tilde{\Psi} = e^{i \alpha (x)} \Psi \;, \;\;\; \bar{\Psi} \to \bar{\tilde\Psi} = e^{ - i \alpha (x)} \bar{\Psi} \;\;\;
{\rm and} \;\;\; A_a \to {\tilde A}_a = A_a - i \alpha_a \;,
\end{equation}
where $A_a$ is the field that needs to be introduced to adsorb the derivative of the gauge parameter $\partial_a \alpha \equiv \alpha_a$. The invariance means that ${\cal A} (\eta_{a b}, {\tilde \Psi} , {\tilde A}_a ) =  {\cal A} (\eta_{a b}, \Psi , A_a )$, i.e. we can use the fields $\Psi$ and $A_a$ or the gauge transformed ones $\tilde{\Psi}$ and ${\tilde A}_a$ and we would not be able to see any difference because the physical effects of a nonzero $\alpha$ are null for this form of the action. For this part the analogy with the Weyl symmetry we just discussed is total: we can use the fields $\Psi$ and $\sigma$ or the Weyl transformed ones $\Psi'$ and $\sigma'$ and we would not be able to see any difference. Furthermore, for the standard gauge theory (\ref{diracgauge}) there is no physical meaning we can ascribe to $\alpha$: it is just the parameter of a kind of rotation of the field point by point. This is the reason why, besides topological non-triviality, the gauge field itself is not observable. Here the analogy with the Weyl case breaks down twice, once because $\sigma$ is directly related to the metric hence indeed it has a physical meaning on its own, and once because, for the case in point of an intrinsically Weyl-gauged action, the Weyl-gauge field is absent hence it is not observable because it is not there in the first place. Thus we can indeed take the view that the non-transformed field $\Psi$ is performing a curved space physics in a metric $e^{- 2 \sigma} \eta$, but identical physical results are obtained by considering the Weyl-transformed field $\Psi '$ performing a flat space physics.

Weyl symmetry is an internal symmetry hence, as such, the associated Noether current is of the form $J_a = \Pi^i_a \delta^W \Phi_i$, where, as usual, ${\Pi^i}^a \equiv \delta {\cal A} / \delta \partial_a \Phi_i$. Now we have to decide how to treat $\sigma$ in (\ref{corerelation1}), namely if we need to include it in the current as one of the fields $\Phi_i$ or not. To properly treat $\sigma$ as a dynamical field we would need to include a kinetic term for it, just like we do for standard gauge theories where we {\it add} to the {\it gauged} action a further term, usually of the form $F_{a b} F^{a b}$, where $F_{a b}$ is the field strength. But the Dirac action in point needs not be Weyl-gauged, that is why there is no $W_\mu$ field, hence the associated kinetic term, if any, can only be relative to the geometric quantity $\sigma$, hence to the metric.

Thus, if no dynamical meaning can be ascribed to $\sigma(x)$, i.e. $\sigma(x)$ is an external field, the Weyl current is
\begin{equation}\label{weylcurrent}
    {\cal J}_a = i e^{(1-n) \sigma} \bar{\Psi} \gamma_a \Psi = i \bar{\Psi}' \gamma_a \Psi' \;.
\end{equation}
The action (\ref{corerelation1}) is also invariant under standard {\it rigid} gauge transformations, i.e. (\ref{standardgaugelocal}) with $\alpha_a = 0$, whose associated Noether current is
\begin{equation}\label{gaugecurrent}
    j_a = e^{(1-n) \sigma} \bar{\Psi} \gamma_a \Psi = \bar{\Psi}' \gamma_a \Psi' = - i {\cal J}_a \;.
\end{equation}
The way we interpret this result is that we have to choose which current to retain as physical\footnote{One could also take a less conservative view and consider complex currents ${\cal J}_a = |{\cal J}_a| e^{i \phi}$ and $j_a = |j_a| e^{i \chi}$, with $|{\cal J}_a| = |j_a|$ and $\chi - \pi / 2 = \phi$, but we shall not do it here.}, and we may as well choose $j_a$. With this choice no further conserved current is introduced when {\it rigid} Weyl symmetry is present. The action is not invariant under local gauge transformation, $\alpha (x)$ but it is under local Weyl and the physical meaning of this symmetry we have explained earlier.

It might look surprising that the Weyl current ${\cal J}_a$ has nothing to do with the energy-momentum tensor but this is as it should be. The Weyl transformations are abelian internal transformations hence no $T_{\mu \nu}$ (a current associated to the non-abelian spatiotemporal group $SO(n,2)$) could ever appear this way. The way to obtain the latter is through
\begin{equation}\label{tmunu}
T^{\mu \nu} \equiv - \frac{2}{\sqrt{g}} \frac{\delta {\cal A}}{\delta g_{\mu \nu}} = i \bar{\Psi} \gamma^{(\mu} \nabla^{\nu)} \Psi - i g^{\mu \nu} \bar{\Psi} \gamma^\lambda \nabla_\lambda \Psi \;,
\end{equation}
whose trace is zero on-shell ($\not\!\nabla \Psi = 0$) with no needs to add improvement terms. The absence of such improvement terms is the effect on $T^{\mu \nu}$ we have to expect from local Weyl symmetry.

\section{Weyl-Gauge Symmetry of Graphene}
\setcounter{equation}{0}

Graphene is a two-dimensional honeycomb lattice of carbon atoms arranged in two triangular sub-lattices, say $L_A$ and $L_B$, whose electrons in the $\pi$-bonds belonging to one sublattice can hop to the nearest neighbor sites of the other sublattice. The electronic properties of graphene are ascribed to these electrons. The elastic properties, instead, are ascribed to the $\sigma$-bonds and involve an energy orders of magnitude stronger than that relative to the $\pi$-bonds. Let us focus on the electronic properties which, in the low-energy approximation, are described by the Hamiltonian (we set $\hbar = 1$)
\begin{equation}\label{HamilGraphene1}
    H = - t \sum_{\vec{r} \in L_A} \sum_{i =1}^3 \left( a^\dagger (\vec{r}) b(\vec{r} +\vec{s}_i)
    + b^\dagger (\vec{r} +\vec{s}_i) a (\vec{r}) \right) \;,
\end{equation}
where $t \simeq 2.7$~eV is the nearest neighbors hopping parameter \cite{katsnelson1} (the next-to-nearest neighbors hopping parameter $t'$ is taken to be zero, and that is the low-energy approximation), $a$ ($a^\dagger$) and $b$ ($b^\dagger$) are anti-commuting annihilation (creation) operators for an electron in the $L_A$ sub-lattice and in the $L_B$ sub-lattice, respectively, and all vectors are two-dimensional $\vec{r} = (x,y)$ and $\vec{s}_i = (s_i^x, s_i^y)$, as described in Fig.~1. The relevant anticommutation relations are
\begin{equation}\label{anticomm}
[a(\vec{r}), a^\dagger (\vec{r'})]_+ = \delta^2 (\vec{r} - \vec{r'}) \;,
[b(\vec{r} + \vec{s}_i), b^\dagger (\vec{r'}+\vec{s}_i)]_+ = \delta^2 (\vec{r} - \vec{r'}) \;,
[a,b]_+ = [a,b^\dagger]_+ = 0
\end{equation}

\begin{figure}
 \centering
  \includegraphics[height=.3\textheight]{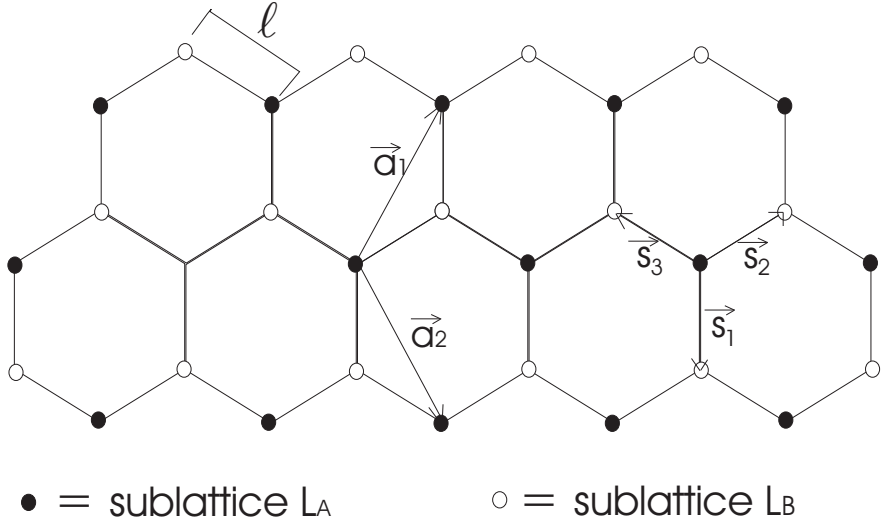}
  \caption{The honeycomb graphene lattice.}
\label{honeycomb}
\end{figure}

First one notices that due to the geometry of the hexagonal lattice the basis vectors of the sublattice $L_A$ are
\begin{equation}\label{avectors}
    \vec{a}_1 = \frac{\ell}{2} (\sqrt{3}, 3) \quad {\rm and} \quad  \vec{a}_2 = \frac{\ell}{2} (\sqrt{3}, - 3) \;,
\end{equation}
and the vectors moving from a site of $L_A$ to the three nearest neighbors $L_B$ sites are
\begin{equation}\label{svectors}
   \vec{s}_1 = \ell (0, - 1) , \quad \vec{s}_2 = \frac{\ell}{2} (\sqrt{3}, 1) \quad {\rm and} \quad  \vec{s}_3 = \frac{\ell}{2} (-\sqrt{3}, 1) \;.
\end{equation}
In the following we shall set the lattice spacing $\ell$ to 1. Taking the Fourier transform, $a(\vec{r}) = \sum_{\vec{k}} a(\vec{k}) e^{i \vec{k} \cdot \vec{r}}$, etc, the Hamiltonian (\ref{HamilGraphene1}) becomes
\begin{equation}\label{HamilGraphene2}
    H = \sum_{\vec{k}} ( f(\vec{k}) a^\dagger (\vec{k}) b(\vec{k}) + {\rm h.c.} )\;,
\end{equation}
where
\begin{equation}\label{fk}
    f(\vec{k}) = - t e^{-i k_y} \left( 1 + 2 e^{i \frac{3}{2} k_x} \cos(\frac{\sqrt{3}}{2} k_x) \right) \;.
\end{equation}
The 1-particle spectrum being given by
\begin{equation} \label{linearep}
E(\vec{k}) = \pm |f (\vec{k})|
\end{equation}
the modes with zero 1-particle energy are easily found as solutions of $f(\vec{k}) = 0$, i.e. $\vec{k}^D_\pm = (\pm \frac{4 \pi}{3 \sqrt{3}}, 0)$, where the superscript $D$ stands for $Dirac$ (points). If we linearize around $\vec{k}^D_\pm$, $\vec{k}_\pm \simeq \vec{k}^D_\pm + \vec{p}$, and $f_\pm (\vec{p}) \equiv f (\vec{k}_\pm) = \pm \frac{3 t}{2} (p_x \pm i p_y)$, $a_\pm (\vec{p}) \equiv a (\vec{k}_\pm)$,  $b_\pm (\vec{p}) \equiv b (\vec{k}_\pm)$, then $H$ in (\ref{HamilGraphene1}) can be written as
\begin{eqnarray}
    H|_{\vec{k}_\pm} & \simeq & \sum_{\vec{p}} [ f_+ a_+^\dagger b_+ +
f_- a_-^\dagger b_- + f^*_+ b_+^\dagger a_+ + f^*_- b_-^\dagger a_- ] (\vec{p}) \nonumber \\
    & = & v_F  \sum_{\vec{p}} [ (a_+^\dagger , b^\dagger_+) \left(\begin{array}{cc} 0 & p_x + i p_y \\ p_x - i p_y & 0 \\ \end{array} \right)
\left(\begin{array}{c} a_+ \\ b_+ \end{array}\right) \nonumber \\
& - & (b_-^\dagger , a^\dagger_-) \left(\begin{array}{cc} 0 & p_x + i p_y \\ p_x - i p_y & 0 \\ \end{array} \right)
\left(\begin{array}{c} b_- \\ a_- \end{array}\right) ] (\vec{p}) \label{HamilGraphene4} \;,
\end{eqnarray}
where $v_F = \frac{3 t}{2}$ is the Fermi velocity (at $\ell = 1$) that we shall now set to 1. If we define
\begin{equation}\label{psi2d}
    \psi_+ \equiv \left( \begin{array}{c} a_+ \\ b_+ \\ \end{array}\right)
\;\; {\rm and} \;\;
\psi_- \equiv \left( \begin{array}{c} b_-\\ a_- \\ \end{array} \right) \;,
\end{equation}
and $\vec{\sigma} \equiv (\sigma_1, \sigma_2)$, with
\begin{equation}\label{sigma12}
    \sigma_1 = \left(\begin{array}{cc} 0 & 1 \\ 1 & 0 \\ \end{array} \right) \; , \;
    \sigma_2 = \left(\begin{array}{cc} 0 & -i \\ i & 0 \\ \end{array} \right) \;,
\end{equation}
the usual Pauli matrices, and we Fourier-transform back to configuration space, then the Hamiltonian (\ref{HamilGraphene4})
becomes
\begin{eqnarray}
    H =  \sum_{\vec{p}} \left(\psi_+^\dagger \vec{\sigma} \cdot \vec{p} \; \psi_+
    - \psi_-^\dagger \vec{\sigma} \cdot \vec{p} \; \psi_- \right) =  - i \int d^2 x \left( \psi_+^\dagger \vec{\sigma} \cdot \vec{\partial} \; \psi_+
    - \psi_-^\dagger \vec{\sigma} \cdot \vec{\partial} \; \psi_- \right) \label{HamilGraphene5} \;.
\end{eqnarray}
The spin $1/2$ structure here is entirely due to the lattice, that is why it is often called pseudospin, but we shall treat it as a proper spin, as suggested, e.g., in \cite{emergentspin}. With the introduction of the four component Dirac spinor\footnote{We prefer to work with four components Dirac spinors because we consider the graphene system as living in a 2 + 1 dimensional spacetime and Weyl spinors, although natural in the two (spatial) dimensions we have considered till now, do not exist in three (spatiotemporal) dimensions.}
\begin{equation}\label{diracspinor}
   \Psi \equiv \left( \begin{array}{c} \psi_+ \\ \psi_- \\ \end{array} \right)
= \left( \begin{array}{c} a_+ \\ b_+ \\ b_- \\ a_- \\ \end{array} \right) \;,
\end{equation}
of the following form for the Dirac matrices
\begin{equation}\label{diracmatrices}
\alpha^i = \left(\begin{array}{cc} \sigma^i & 0 \\ 0 & - \sigma^i \\ \end{array} \right) \;, \;
\beta = \left(\begin{array}{cc} 0 & I_2 \\ I_2 & 0 \\ \end{array} \right) \;, \; i = 1, 2 \;,
\end{equation}
which lead to the usual definition of $\gamma$-matrices
\begin{equation}\label{gammamatrices}
    \gamma^0 \equiv \beta \;, \; \gamma^i \equiv \beta \alpha^i \;,
\end{equation}
satisfying
\begin{equation}\label{anticommgamma}
    [ \gamma^a , \gamma^b ]_+ = 2 \eta^{a b} I_4 \;,
\end{equation}
with $a,b = 0,1,2$, the ``$v_F$ relativistic'' Dirac Hamiltonian in (\ref{HamilGraphene5}) can be further compacted to
\begin{equation}\label{HamilGraphene6}
    H = \int d^2 x \left( - i \bar{\Psi} \; \vec{\gamma} \cdot \vec{\partial} \; \Psi \right)
    \equiv \int d^2 x {\cal H}  \;,
\end{equation}
where $\bar{\Psi} = \Psi^\dagger \gamma^0$.

Thus, starting from the instance that the energy dispersion relations (\ref{linearep}) become linear at the Dirac points, taking into account only the low energy contributions, including the effect of the honeycomb structure of Fig.~1 and considering situations that do not include defects or impurities, the effective description of the electronic properties of graphene is given by a special relativistic-like Hamiltonian from which
through customary Legendre transformation, ${\cal L} = i \Psi^\dagger \; \dot{\Psi} - {\cal H}$, a Lagrangian density can be derived. What is obtained  is precisely the scale- and conformal-symmetric massless Dirac action in flat space we discussed in the previous Section with $n=3$
\begin{equation}\label{actiongrapheneflat}
    A = i \int d^3 x \bar{\Psi} \; \gamma^a \partial_a \; \Psi \;.
\end{equation}
Now, let us suppose that the graphene sheet is {\it bent}. By this we mean that the two-dimensional sheet of graphene is not planar, hence curvilinear coordinates are better suited. A typical situation we have in mind is, of course, that of the ripples \cite{ripples1}, \cite{ripples2}, \cite{ripples3} (see also the review \cite{review2010}), that are found in experiments and that are the subject of much theoretical and experimental investigations on their origins and their impact on the electronic properties of graphene, but other geometries are also within the reach of the following analysis. The graphene action to consider is then
\begin{equation}\label{actiongraphenecurve}
    {\cal A} = i \int d^3 x \sqrt{g} \; \bar{\Psi} \gamma^a E^\mu_a (\partial_\mu + \frac{1}{2} \omega_\mu^{\; b c} J_{b c}) \Psi \;,
\end{equation}
where $g_{\mu \nu}$ is a three-dimensional metric that bears the information about the bending in the two spatial dimensions (we shall be more precise about this in the next Section). Thus the graphene action, in the limits recalled above, is a particular instance of the ``intrinsically Weyl-gauged'' action for massless Dirac fields we have discussed before.

{\it Every bending} of the two-dimensional sheet produces a conformally flat two-dimensional metric $g_{\alpha \beta}$, $\alpha, \beta = 1,2$ for the simple reason that every two-dimensional metric is conformally flat. It is matter of finding the coordinates system where this is evident, but for a well known general argument this is always possible. These conformally flat $g_{\alpha \beta}$s describe graphene sheets of which: (i) one is planar, hence intrinsically (and extrinsically) flat; (ii) some are non-planar but are in fact intrinsically flat; (iii) some have constant intrinsic curvature; (iv) some have non-constant intrinsic curvature. Due to the local Weyl invariance in any dimension that we discussed in the previous Section, if the dimensionality of the problem was indeed two we would have immediately the result that in each one of the cases (i)-(iv) the electronic properties of graphene are described by a conformally trivial system and we could take advantage from this every time.

On the other hand, the correct dimensionality is three because there is time. In the next Section we shall see that the presence of time, the special relativistic-like form of the action and the two-dimensional conformal flatness of the graphene sheet naturally lead to a general relativistic-like metric, namely a 2+1 dimensional conformally flat metric, including the truly curved. We resort to such a discovery while asking the question: is there a particular shape of the two-dimensional sheet that gives raise to a conformally trivial Dirac system also in \textit{three} dimensions, i.e. including time? We postpone the answer to the next Section as we want to come back now to the effect of local Weyl invariance on the density of states of graphene.

For graphene, the quantum version of the local Weyl invariance discussed in general in the previous Section can have striking consequences. First we notice that the electronic LDOS can be written in terms of the two-point function \cite{rickayzen} (see also \cite{dnapaper} and \cite{vozmediano})
\begin{equation}\label{LDOSflat}
    \rho (E, \vec{x}) \equiv \frac{1}{\pi} {\rm Im Tr} \left( G (E,\vec{x},\vec{x}) \gamma^0 \right) \;,
\end{equation}
where $G$ stands for a generic Green function ($S$, $S'$, $S^\sigma$ or $S^{-\sigma}$ of the previous Section) and to obtain its energy dependence it is customary to move to the Hamiltonian form of the Dirac propagator, formally: $G(E) = 1/(H - E)$. Then we need to apply the general discussion to the case of graphene, by first asking whether it makes sense to have a truly 2+1 dimensional conformally flat situation for graphene. As said we shall face this problem in detail in the following Section but we anticipate that the answer is ``yes''. Thus, what is left to do is to identify the metric and define the corresponding measuring procedure such that
\begin{equation}\label{LDOScurved}
    \rho' (E, \vec{x}) =  \frac{1}{\pi} {\rm Im Tr} \left( S' (E,\vec{x},\vec{x}) \gamma^0 \right) =  \frac{1}{\pi} {\rm Im Tr} \left( S (E,\vec{x},\vec{x}) \gamma^0 \right) = \rho (E, \vec{x}) \;,
\end{equation}
and the LDOS for the planar sheet is the same as that for a curved sheet when the two-dimensional curvature is such that the 2+1 dimensional metric is conformally flat.

\section{The ``nearly two-dimensional'' {\it Ansatz} for the metric}
\setcounter{equation}{0}

The two-dimensional sheet in immersed in a three-dimensional space $(x,y,z)$. One can easily set the frame so that the profile of the sheet is $z(x,y)$. Including time then we have the extrinsic frame $\beta^a = (t,x,y,z(x,y))$ and the intrinsic frame  $\alpha^\mu = (t,x,y)$, where flat ($a = 0,1,2,3$) and curved ($\mu = 0,1,2$) indices need to be noticed. The (standard) induced metric procedure gives for the 2 + 1 dimensional system the metric
\begin{equation}\label{inducedg}
    g^{(3)}_{\mu \nu} = \eta_{a b} \frac{\partial \beta^a}{\partial \alpha^\mu} \frac{\partial \beta^b}{\partial \alpha^\nu} \;,
\end{equation}
where the choice of four dimensional flat metric, $\eta_{a b}$, is dictated by the special relativistic-like form of the action, we just need to remember that the limiting speed for the electronic system is $v_F$ and not $c$, all the rest goes through. With this choice for $\beta^a$ and $\alpha^\mu$ the electrons on graphene see a {\it spacetime} metric of the form
\begin{equation}\label{nearly2d}
    g^{(3)}_{\mu \nu} (x,y) = \left(\begin{array}{cc} 1 & 0 \\ 0 & g^{(2)}_{\alpha \beta} (x,y) \\ \end{array} \right) \;,
\end{equation}
where $\alpha, \beta = 1,2$ are the spatial two-dimensional indices and $g^{(2)}_{1 1} = - 1 - (\partial z / \partial x)^2$, $g^{(2)}_{2 2} = - 1 - (\partial z / \partial y)^2$, $g^{(2)}_{1 2} = g^{(2)}_{2 1} = - (\partial z / \partial x)(\partial z / \partial y)$. This we call the ``nearly two-dimensional'' {\it Ansatz} because, although the metric $g^{(3)}_{\mu \nu}$ is indeed three-dimensional, the curvature content appears to be all in the two-dimensional spatial part. To see it one first notices that the scalar curvatures for $g^{(3)}_{\mu \nu}$ and $g^{(2)}_{\alpha \beta}$ are equal
\begin{equation}\label{r3=r2}
    R^{(3)} = R^{(2)} \;,
\end{equation}
and then that the Ricci tensors for the two metrics are such that
\begin{equation}\label{ricci3=ricci2}
{R^{(3)}}_{\mu \nu} =   \left(\begin{array}{cc} 1 & 0 \\ 0 & {R^{(2)}}_{\alpha \beta} \\ \end{array} \right)\;.
\end{equation}
These results are due to the {\it Ansatz} (\ref{nearly2d}). On the other hand in any three-dimensional space the Riemann tensor (the tensor that bears the full information on curvature) ${R^{(3)\; \rho}}_{\sigma \mu \nu}$ is proportional to $R^{(3)}$ and ${R^{(3)}}_{\mu \nu}$ as follows (see, e.g., \cite{thooft}, \cite{djt}, \cite{carlip})
\begin{equation}\label{riemannthree}
{{R^{(3)}}^{\; \rho \sigma}}_{\mu \nu} = \epsilon^{\rho \sigma \lambda} \epsilon_{\mu \nu \kappa} ({R^{(3)}}_\lambda^\kappa
- \frac{1}{2} \delta_\lambda^\kappa R^{(3)}) \;,
\end{equation}
while in any two-dimensional space (see the Appendix)
\begin{equation} \label{ricci2}
R^{(2)}_{\alpha \beta} = \frac{1}{2} g^{(2)}_{\alpha \beta} R^{(2)} \;.
\end{equation}
Thus, from the results (\ref{r3=r2})-(\ref{ricci2}) we see that the whole information on the curvature of the three-dimensional spacetime with metric (\ref{nearly2d}) is actually in the two-dimensional scalar curvature $R^{(2)}$, i.e. the Riemann tensor only has one independent component, a fact that characterizes two-dimensions, for us the $(x,y)$ space of the sheet. This, apparently, means that everything related to curvature is happening in the two-dimensional spatial part and we have built on purpose such a situation to be in the most conservative position: (i) no curvature on the time part, (ii) the metrics only depend on $x$ and $y$.

\begin{figure}
 \centering
  \includegraphics[height=.3\textheight]{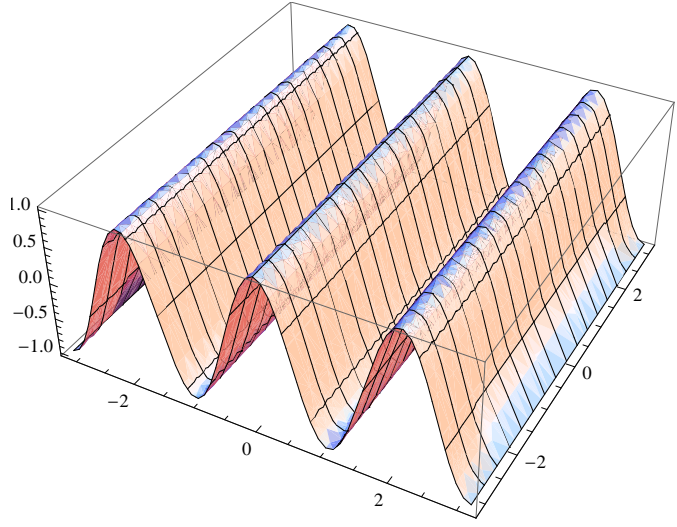}
  \caption{An intrinsically flat bending corresponding to the profile $z = \cos(3 x)$. This profile gives raise to a flat 2+1 dimensional metric in the appropriate coordinates.}
\label{3DACosnX}
\end{figure}

Our goal is to see whether a conformally flat $g^{(3)}_{\mu \nu}$ makes sense. Namely, we want to see whether by imposing the conditions for three-dimensional conformal flatness we obtain a reasonable condition for the shape of the two-dimensional graphene sheet. If that happens then we could reverse the argument and say that for such a shape the three-dimensional metric is indeed conformally flat, including the time part of it, hence we would be in a truly general relativistic-like system, although conformally trivial. What could happen (and we were expecting to happen!) is that only flat sheets would be allowed. This case is also of interest, as we shall show, because we are focusing on the intrinsic geometric properties, hence, e.g., a shape like that in Fig.~\ref{3DACosnX} is flat. On the other hand if flat two-dimensional sheets were the only solution then the only conformally flat $g^{(3)}_{\mu \nu}$ would be the flat metric $\eta^{(3)}_{\mu \nu}$.

Our problem becomes that of solving the third order partial differential equations
\begin{equation}\label{cotton1}
    C_{\mu \nu} = 0 \;,
\end{equation}
where
\begin{equation}\label{cotton2}
    C_{\mu \nu} = \epsilon_{\mu \lambda \kappa} \nabla^\lambda {R^{(3)}}^\kappa_\nu + \epsilon_{\nu \lambda \kappa} \nabla^\lambda {R^{(3)}}^\kappa_\mu \;,
\end{equation}
is (proportional to) the so-called Cotton tensor \cite{djt} \cite{cs3} and its vanishing is necessary and sufficient condition for a three-dimensional metric to be conformally flat. In four dimensions this would not be true as it is the vanishing of the Weyl tensor that gives the necessary and sufficient condition for the conformal flatness. The latter tensor is identically zero in three dimensions, and this is why the Riemann tensor takes the form (\ref{riemannthree}).

\begin{figure}
 \centering
  \includegraphics[height=.3\textheight]{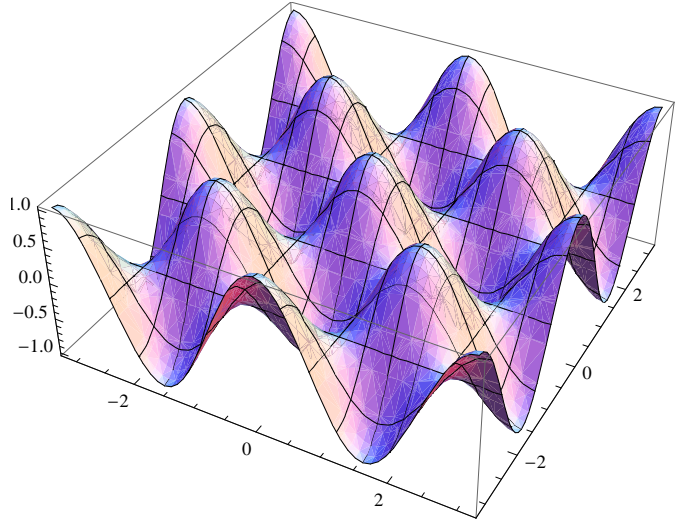}
  \caption{An ``egg-tray'' shaped (intrinsically curved) sheet corresponding to the profile (\ref{zprofile}) with $C=1$ and $n=m=2$. We cannot tell whether there is a choice of $C, n, m$ such that this kind of profiles give raise to a conformally flat 2+1 dimensional metric.}
\label{3DACosnXCosmY}
\end{figure}

The task to solve the partial differential equations (\ref{cotton1}) may become very difficult, hence we turned to the help of the package for Mathematica in \cite{rgtc}. With that we have tried with the general profile
\begin{equation}\label{zprofile}
    z(x,y) = C \cos(n x) \cos(m y) \;,
\end{equation}
with $n,m$ integers (see Fig.~\ref{3DACosnXCosmY}) that could well simulate the ripples found on suspended graphene \cite{ripples1}, in the hope to find some values of the amplitudes $C$ and/or of the frequencies $n, m$ for which (\ref{cotton1}) has solutions, besides the flat solutions $n = 0, m = {\rm any}$ and $n = {\rm any}, m = 0$.
The task revealed itself to be beyond the capabilities of the program and/or of the machine. Thus we resorted to approach the problem using the general argument that, for any given profile $z(x,y)$, there exists a coordinate system on the sheet, say it $(\tilde{x}, \tilde{y})$, such that $g^{(2)}_{\alpha \beta} = - e^{-2 \sigma (\tilde{x}, \tilde{y})} \delta_{\alpha \beta}$. To find such coordinate transformation is surely possible in principle but could be tricky in practice (see, e.g., \cite{funnykink} for some examples in a 2+1 dimensional case), but the differential equations (\ref{cotton1}) surely become much more tractable. First, in the coordinates $(\tilde{x}, \tilde{y})$ the metric (\ref{nearly2d}) becomes
\begin{equation}\label{CFnearly2d}
    g^{(3)}_{\mu \nu} (\tilde{x}, \tilde{y}) = \left(
                    \begin{array}{ccc}
                      1 & 0 & 0 \\
                      0 & - e^{-2 \sigma (\tilde{x}, \tilde{y})} &  0 \\
                      0 & 0 & - e^{-2 \sigma (\tilde{x}, \tilde{y})} \\
                    \end{array}
                  \right)
 \;.
\end{equation}
With this equations (\ref{cotton1}) become
\begin{eqnarray}
\partial_{\tilde{x}} \triangle \sigma + 2 (\partial_{\tilde{x}} \sigma) \triangle \sigma & = & 0 \label{main1} \\
\partial_{\tilde{y}} \triangle \sigma + 2 (\partial_{\tilde{y}} \sigma) \triangle \sigma & = &  0 \label{main2} \;,
\end{eqnarray}
where $\triangle = \partial_{\tilde{x}}^2 + \partial_{\tilde{y}}^2$ is the flat space two-dimensional Laplace operator. Now we compute $R^{(2)}$ in these coordinates and find (see the Appendix)
\begin{equation}\label{Rsigma}
    R^{(2)} = 2 e^{2 \sigma} \triangle \sigma \;.
\end{equation}
By using (\ref{Rsigma}) in (\ref{main1}) and (\ref{main2}) we obtain that the request for $g^{(3)}_{\mu \nu}$ in (\ref{CFnearly2d}) to be conformally flat amounts to have a two-dimensional sheet with constant intrinsic curvature $R^{(2)}$. This, in terms of the two-dimensional conformal factor $\sigma$, amounts to have
\begin{equation}\label{liouville}
    \triangle \sigma = K e^{-2 \sigma} \;,
\end{equation}
with {\it constant} Gaussian curvature $K = R^{(2)} / 2$. We wrote the condition on the $\sigma$ in this fashion because this way it is easy to recognize it as a classic equation of the geometry of two-dimensional surfaces \cite{geometry1}, \cite{geometry2}, namely \textit{Liouville's equation} \cite{liouville1}. Among the advantages of having such a classic equation as our constraint is that all solutions are known \cite{liouville1}, see also \cite{liouville2}-\cite{liouville4}.

Thus we have found that \textit{for any graphene sheet that is truly curved, but with constant $K$, the electrons see a truly curved, but conformally flat, three-dimensional spacetime}. To explicitly see it one needs to find a third coordinate frame, now involving time (as in principle we cannot and should not exclude it) such that
\begin{equation}\label{3DCFnearly2d}
    g^{(3)}_{\mu \nu} (T,X,Y) = e^{-2 \Sigma (T,X,Y)} \left(
                    \begin{array}{ccc}
                      1 & 0 & 0 \\
                      0 & - 1 & 0 \\
                      0 & 0 & - 1 \\
                    \end{array}
                  \right)
 \;,
\end{equation}
with $\Sigma = \sigma$ only in the trivial case of constant $\sigma$ (and $T = e^{-\sigma} t, X = \tilde{x}, Y = \tilde{y}$).

When the two-dimensional spatial metric is such that the Liouville equation (\ref{liouville}) is satisfied, we have the ``most gentle'' general relativistic-like situation we could think of, namely the conformally trivial one. So, while on the one hand, this means that many characteristics of the flat case are untouched in the conformally trivial case (as we explained before), on the other hand {\it some} features will be general relativistic-like and this, in turn, opens the doors to speculate whether more general settings of this kind are meaningful for the physics of electrons on graphene. On the latter (``exotic'') scenario we will briefly comment in the last Section. Let us now look more closely at the rich landscape that the constraint (\ref{liouville}) opens-up.

When $K = 0$ we are in the particular case of flat sheets, that clearly gives flat three-dimensional metrics. In this case the $\sigma$s permitted are all the infinite solutions of the Laplace equation
\begin{equation}\label{laplace}
    \triangle \sigma = 0 \;,
\end{equation}
that, as well known, are the harmonic functions in two dimensions. It is also well known that in the complex domain, $w = \tilde{x} + i \tilde{y}$ and
$\sigma (w) = \sigma_R (\tilde{x},\tilde{y}) + i \sigma_I (\tilde{x},\tilde{y})$, when $\sigma_R$ and $\sigma_I$ obey the Cauchy-Riemann conditions for analyticity, $\partial_{\tilde{x}} \sigma_R = \partial_{\tilde{y}} \sigma_I$ and $\partial_{\tilde{x}} \sigma_I = - \partial_{\tilde{y}} \sigma_R$, then they are also harmonic functions. This means that that are infinite ways we can bend the graphene sheet by keeping it intrinsically flat, the representative of one possible family is given in Fig.~\ref{3DACosnX}, and that this infinite family has tight links with the infinite dimensional Virasoro algebra, the algebra of the conformal group in two dimensions, with generators $L_n = - w^{n+1} \partial_w$, even if we are dealing with a three-dimensional spacetime. On these matters see, e.g., \cite{cft}. Of course, to bend the graphene sheet it costs elastic energy even though the shape reached is intrinsically flat. Thus, from the point of view of the elastic energy, there is no such a thing as a family of invariant shapes. What is invariant instead is the way the electrons behave within the intrinsically flat sheets: for them there is no distinction among members of the class.

If we are able to find the coordinate transformation $(t, \tilde{x}, \tilde{y}) \to (T,X,Y)$ that gives the relation between the  $\sigma$ of the two-dimensional metric (\ref{CFnearly2d}) and the  $\Sigma$ of the three-dimensional metric (\ref{3DCFnearly2d}) we would explicitly see how the Virasoro algebra, that surely is there for $\sigma$, enters the three-dimensions via $\Sigma$. This is an interesting mathematical problem on its own and deserves more studies \cite{3dvirasoro}.

Another interesting point is relative to true curvature for a honeycomb lattice structure as the one we have in mind here. As well known, see, e.g., \cite{KatVol92}, \cite{Kleinert}, \cite{vozmedianodefect} and also \cite{cejb}, for such a lattice the curvature can only enter via the appearance of a defect, a fivefold one to have a positive $K$, a sevenfold one to have a negative $K$. We have not taken this into account in this paper (but shall face this in follow-up work \cite{TMGmarket}, \cite{TMGdynamical}) hence, in this respect, we are considering the graphene sheet as a continuum. Nonetheless, we want to notice here that if a ``no-defect condition'' is imposed on true graphene then there is no way out to have flat sheets and this, in turn, is not as a limiting situation as one might think because the number of allowed shapes is still infinite (and related to the Virasoro algebra).

Thus, within our approximation of curvature not due to defects, the next case to consider is the constant $K$. If we once more move to the complex domain equation (\ref{liouville}) becomes
\begin{equation}\label{liouvillecomplex}
    \partial_w \partial_{\bar{w}} \ln \varphi = - \frac{K}{2} \varphi \;,
\end{equation}
with $ 4 \partial_w \partial_{\bar{w}} = \partial_{\tilde{x}}^2 + \partial_{\tilde{y}}^2$ and $\varphi = e^{-2\sigma}$. The general solutions are \cite{liouville1}
\begin{equation}\label{solliouv}
    \varphi (w) = \frac{4}{K} \frac{|f'(w)|^2}{(1+|f(w)|^2)^2} \;,
\end{equation}
with $f(w)$ any analytic function, hence, again, an infinite family of solutions like for the harmonic functions. In this case as well this typically two-dimensional rich structure is enjoyed by a three-dimensional system but to clearly see how this is reflected on the three-dimensional $\Sigma$ one needs to find the coordinate transformations \cite{3dvirasoro} $(t, \tilde{x}, \tilde{y}) \to (T,X,Y)$. Nonetheless, this intriguing structure underlays our problem and should be explored carefully in relation to possible applications to graphene \cite{LDOSreal}. For instance, vortices have been studied in the context of graphene from various perspectives, see, e.g., \cite{jackiwpi}, \cite{stone}, and we can add to those a new one by considering that among the solutions (\ref{solliouv}) there are (non-topological) vortex solutions \cite{liouville4} given by $f(w) = w^{-n}$, with $n$ positive integer to which corresponds
\begin{equation}\label{solliouvvortex}
    \varphi (w) = \frac{4 n^2}{K} \frac{r^{-2(n+1)}}{(1+r^{-2n})^2} \;,
\end{equation}
where $r = |w|$. Current work is devoted to find the actual profile $z(x,y)$ corresponding to those solutions and the impact on the density of states \cite{LDOSreal}.

\section{Conclusions and Discussion}
\setcounter{equation}{0}

We can then conclude that, within the approximations explained throughout the paper\footnote{Besides the low energy limit and the $v_F$-like special relativistic behavior, one crucial approximation here is that curvature is not introduced via disclination defects, hence, in this respect we are treating the lattice as a continuum.}, when the graphene sheet has a constant Gaussian curvature the three-dimensional spacetime the electrons move in is conformally flat and truly curved in all directions, including time, i.e. they experience a truly general relativistic environment. Nonetheless, due to the local Weyl symmetry enjoyed by the massless Dirac theory, these effects cannot be seen classically nor quantum-mechanically, unless ``non-invariant measurements'' are performed. Geometrically, by ``invariant measurements'' we mean processes of measurement for which times and lengths are given by $ds^2 = e^{-2 \Sigma} (dt^2 - d{\vec x}^{\; 2})$, hence the significant vacuum expectations are of the kind $\langle 0| O |0 \rangle$ and, due to the invariance, they are equal to the flat vacuum expectations $\; '\langle 0| O' |0 \rangle'$, where $O$ is a generic operator and we use the notation of the last Section as for conformal factors. Physically, the processes of invariant measurements are related to the kind of spacetime $\Sigma$ entails. One consequence of the Weyl symmetry leading to this general relativistic behaviors is that the electronic LDOS is invariant under the process of curving the graphene sheet, provided, as said, the measurements are of the invariant kind and provided the two-dimensional curvature is compatible with three-dimensional conformal flatness. For these reasons our conclusions in this respect are not necessarily contradicting the results of \cite{vozmediano} (see also the more recent \cite{raoux}).

Of course, this paper is only a preliminary step towards a match with experiments. First, while the flat spacetime relativistic-like description of (\ref{actiongrapheneflat}) is fully justified from the experimental point of view, the same cannot be said for the curved spacetime description of (\ref{actiongraphenecurve}) and, in general, this is still an open problem. Nevertheless, we have proved here that such curved spacetime description indeed makes sense at least for the (infinite) classes of metrics considered here (which physically correspond to intrinsically flat and constantly intrinsically curved graphene sheets). For those metrics, if one accepts the flat spacetime description then one must accept also a curved spacetime description, simply because, through a Weyl redefinition of the fields, the action is the same. This was the main motivation behind our trial metric (\ref{nearly2d}) in the first place: which metric naturally gives a curved spacetime graphene action that reduces to the experimentally sound flat spacetime graphene action?

Second, we need to find the coordinate transformations that give $\Sigma$ in terms of the two dimensional $\sigma$. Current work is dedicated to this
task \cite{LDOSreal}. There we shall, on the one hand, focus on real graphene configurations that can give raise to a three dimensional conformally flat spacetime, on the other hand, we shall be able to appreciate what kind of spacetime is encoded by such bendings (besides appreciating the lifting to three dimensions of the Virasoro and Liouville structures \cite{3dvirasoro}). In \cite{LDOSreal} we shall also propose dedicated experiments (or at least the use of existing data) to prove (or disprove) at once that the curved space-time action (\ref{actiongraphenecurve}) indeed describes electronic transport on curved graphene sheet and that Weyl-symmetry is truly in place. It is worth noticing here that, although it will surely be very interesting to see how time is affected in the inner reference frame through $g^{(3)}_{00} = \exp(-2 \Sigma)$, we have already proved here that it is only a matter of a coordinate change (not a change of reference frame) to rewrite the metric where $g^{(3)}_{00} = 1$, i.e. (\ref{nearly2d}), as
$\exp(-2 \Sigma) \eta_{\mu \nu}$. The latter coordinate change always exists if the graphene sheet has constant curvature, and indeed it seems a non-impossible task for the experimentalists to realize such sheets (see, e.g., \cite{Terrones2001} for an old overview on generic nanomaterials, and \cite{Kholmanov2009} for a more recent work on graphene).

Once the relevance of conformally flat spacetimes is established, a natural question is whether it makes sense to reverse the logic and consider more general cases of three-dimensional conformally flat spacetimes than those considered here to see which sort of behavior for two-dimensional graphene they correspond to \cite{TMGmarket}. Intriguing examples are the celebrated conformally flat black hole in 2+1 dimensions found by Ba\~{n}ados, Teitelboim and Zanelli  \cite{btz} (BTZ) and the kink spacetime of \cite{cs3}. The first case would represent a black hole in a lab \cite{TMGmarket}, the second case is worth studying on its own right but also to explore whether fractionalization of fermion number could occur in such topologically non-trivial background \cite{jackiwrebbi}, \cite{jackiwsemenov}, \cite{mitthesis}, \cite{TMGmarket}.

The latter examples of three-dimensional conformally flat spacetimes are just two among many solutions of the full theory of gravity in three dimensions that was proposed in the early eighties in \cite{djt} and is still under intense investigation (see \cite{strominger2010} for a recent paper). For that theory the Euler-Lagrange equations are
\begin{equation}\label{TMGequations}
    R_{\mu \nu} - 2 \Lambda g_{\mu \nu} + \frac{1}{\mu} C_{\mu \nu} = 0 \;,
\end{equation}
where all tensors refer to three dimensions and have been defined in the paper (note that this $C_{\mu \nu}$ differs from the one we used in (\ref{cotton2}) only by a factor proportional to $1/\sqrt{g}$, \cite{djt}, \cite{cs3}), $\Lambda$ is the cosmological constant, and $\mu$ is a parameter with the dimensions of a mass. While the first two terms on the left side of (\ref{TMGequations}) descend from a standard Einstein-Hilbert action (hence are the same in any dimension), the third term descends from an action that is nonzero only in three dimensions\footnote{It is actually nonzero when $n=4k+3$, for $k=0,1,2,...$.} and it is called the Chern-Simons gravitational term \cite{cs3}
\begin{equation}\label{cschris}
{\cal A}_{\rm CS} (g_{\mu \nu})= \frac{1}{4 \pi^2} \int d^3 x \epsilon^{\mu \nu
\lambda} \left( \frac{1}{2} \Gamma^\rho_{\mu \sigma} \partial_\nu
\Gamma^\sigma_{\lambda \rho} + \frac{1}{3} \Gamma^\rho_{\mu
\sigma} \Gamma^\sigma_{\nu \tau} \Gamma^\tau_{\lambda \rho}
\right) \;,
\end{equation}
where the Christoffel connection can be expressed in terms of the metric, then variations of ${\cal A}_{\rm CS}$ with respect to $g_{\mu \nu}$ give the Cotton tensor. Hence, conformally flat spacetimes ($C_{\mu \nu} = 0$) are also solutions of the full theory ($R_{\mu \nu} = 2 \Lambda g_{\mu \nu}$). For instance, the kink spacetime mentioned above is a solution of $C_{\mu \nu} = 0$, hence of the full theory, while the BTZ black hole is a solution of $R_{\mu \nu} = 2 \Lambda g_{\mu \nu}$, hence, again, of the full theory and conformally flat. Thus, if we intend to hunt for the Weyl invariant configurations of graphene depicted here, we should restrict our interest to this class of solutions of (\ref{TMGequations}) and answer, case by case, the questions about shapes, invariant measuring procedures, invariance of the density of states, etc. Notice that, in general, $C_{\mu \nu} = 0$ does not give the constraint of Eqs. (\ref{main1}) and (\ref{main2}), hence also \textit{nonconstant} Gaussian curvature of the two-dimensional sheet could give raise to a 2+1 dimensional conformally flat spacetime, hence to a Weyl invariant density of states.

One crucial issue not considered here is about the constraints on the shapes set by the elastic properties of the graphene sheet. For instance, in our model, there is no limitation to the amplitudes of the intrinsically flat $z(x,y) = C \cos(n x)$ (see Fig.~2) besides the obvious limitations arising from the finite size of the sheet (for a square sheet of side $L$, $C_{max} = L/2$ with $n=1$). These inputs can only come from the elastic (and dominant) part of the total energy. One direction to investigate, then, is to construct a gravity/elastic theory of the graphene sheet along the lines of what has been done in \cite{KatVol92} and \cite{Kleinert}, but introducing a further element in that construction that is the elastic counterpart of the Chern-Simons gravitational term (\ref{cschris}) \cite{TMGdynamical}. Of course to this end we need to revisit our analysis because disclination defects are necessary to have a nonzero curvature. Nonetheless, if the defects are few and localized (for instance only 12 five-folded defects are necessary to have a sphere) and the graphene sample is large enough, the analysis presented here should still apply, although less rigorously.

Let us close this paper with a speculation on the nature of ripples \cite{ripples1}-\cite{ripples3}. Since it costs extra energy to create a defect it seems reasonable to ask whether ripples are intrinsically flat. We could exactly answer this question by using the techniques and invariance (classical and quantum) presented here with no approximations because the spacetime is flat: if ripples are intrinsically flat the density of states would be fully invariant, while its non-invariant behavior would be simply due to a coordinate choice effect (not to an observer effect) and, furthermore, the profile $z(x,y)$ would correspond to a two-dimensional conformal factor $\sigma(x,y)$ that is an harmonic function \cite{LDOSreal}.

\bigskip

{\bf \Large Acknowledgments}

\noindent I am indebted to Siddhartha Sen who suggested to me that Weyl symmetry could be relevant for the physics of graphene in the first place, and discussed with me on the density of states and other matters. I thank Petr Jizba, Gaetano Lambiase, Antonello Scardicchio and Martin \v{Z}ofka for useful interactions and Sotirios Bonanos for providing for free a high quality calculation tool for geometric tensors.

\appendixa

\noindent As usual, in $n$ dimension,
\begin{equation}\label{a1}
    [\nabla_\mu , \nabla_\nu]_{-} V_\lambda = {R^\rho}_{\lambda \mu \nu} V_\rho - {T^\rho}_{\mu \nu} \nabla_\rho V_\lambda \;,
\end{equation}
with $\nabla_\mu V_\nu = \partial_\mu V_\nu - {\Gamma^\lambda}_{\mu \nu} V_\lambda$ and the Riemann and torsion tensors are
\begin{equation}\label{a2}
    {R^\rho}_{\lambda \mu \nu} = \partial_{[\nu} {\Gamma^\rho}_{\mu] \lambda} + {\Gamma^\rho}_{[\nu \sigma} {\Gamma^\sigma}_{\mu] \lambda} \;,
    \quad    {T^\rho}_{\mu \nu} = {\Gamma^\rho}_{[\mu \nu]} \;,
\end{equation}
respectively. By using $\Gamma_{\mu \nu}^\lambda = E^\lambda_a ({\delta^a}_b \partial_\mu + {{\omega_\mu}^a}_{b} ) e^b_\lambda$, which descends from $\nabla_\mu e^a_\nu = 0$, as explained in the paper, one has
\begin{equation}\label{a4}
    {(R_{\mu \nu})^a}_b = \left( \partial_{[\mu} \; \omega_{\nu]} + \omega_{[\mu} \; \omega_{\nu]} \right)^a_{\;\; b} \;,
\end{equation}
and
\begin{equation}\label{a5}
    R^\rho_{\; \lambda \mu \nu} = E^\rho_{\;\; a} (R_{\mu \nu})^a_{\; \; b} e^b_{\;\; \lambda} \;, \;\;
    R_{\lambda \nu} = E^\mu_{\;\; a} (R_{\mu \nu})^a_{\; \; b} e^b_\lambda \;, \;\;
    R = E^\mu_{\;\; a} (R_{\mu \nu})^{a b} {E^\nu}_b \;,
\end{equation}
and
\begin{equation}\label{a6}
T^a_{\;\; \mu \nu} = \left( \partial_{[\mu} \; e_{\nu]} + \omega_{[\mu} \; e_{\nu]} \right)^a \;.
\end{equation}
where $T^\lambda_{\;\; \mu \nu} = {E_a}^\lambda T^a_{\;\; \mu \nu}$. For this paper we take $T^a_{\;\; \mu \nu} = 0$ (although we keep track of when we require that to hold) which gives
\begin{equation}\label{a7}
\omega_{a b c}= \frac{1}{2} \left( E^\mu_a E^\nu_c (\partial_\nu
e_{b \mu} - \partial_\mu e_{b \nu}) - E^\mu_c E^\nu_b
(\partial_\nu e_{a \mu} - \partial_\mu e_{a \nu}) + E^\mu_b
E^\nu_a (\partial_\nu e_{c \mu} - \partial_\mu e_{c \nu}) \right) \;,
\end{equation}
where $\omega_{a b c} = E^\mu_a {\omega_\mu}_{b c}$. With this (or by using (\ref{spinconngeneral}) in the paper and ${\Gamma^\lambda}_{\mu \nu}$ in terms of $g_{\mu \nu}$) we have the scaling properties of the spin connection
\begin{equation}\label{a8}
    \Delta {\omega_\mu}^{a b} \equiv {\omega_\mu}^{a b} (e^\sigma {e^a}_\mu) - {\omega_\mu}^{a b} ({e^a}_\mu)
    = {e^{[a}}_\mu{e^{b]}}_\nu \sigma^\nu \;.
\end{equation}

For $n=3$, $\omega_{\mu \; a b} = \epsilon_{a b c} \,\omega_\mu^{\;\; c}$, and $\omega_\mu^{\; a} = e^b_\mu \omega_b^{\; a}$ and
\begin{equation}\label{a9}
\omega_a^{\; d}  = \frac{1}{2} \epsilon^{b c d} \left( e_{\mu a} \partial_b E_c^\mu + e_{\mu b} \partial_a E_c^\mu
+ e_{\mu c} \partial_b E_a^\mu \right) \;,.
\end{equation}

For $n=2$, $\omega_{\alpha \; a b} = \epsilon_{a b} \omega_\alpha$ (or ${\omega_\alpha} = -\frac{1}{2} \epsilon^{a b} \omega_{\alpha \; a b}$) and
\begin{equation}\label{a10}
    \epsilon^{a b} {E_a}^\alpha {E_b}^\beta = \epsilon^{\alpha \beta} \det E = \frac{1}{\sqrt{g}} \epsilon^{\alpha \beta} \;, \;\;
    \epsilon_{a b} {e^a}_\alpha {e^b}_\beta = \epsilon_{\alpha \beta} \det e = \sqrt{g} \epsilon_{\alpha \beta} \;,
\end{equation}
and $\partial_{[\beta} \omega_{\alpha]} = \epsilon_{\beta \alpha} \epsilon^{\gamma \delta} \partial_{\delta} \omega_{\gamma}$ which give
\begin{equation}\label{a11}
    R_{\alpha \beta \gamma \delta} =
    \sqrt{g} \epsilon_{\alpha \beta} \epsilon_{\delta \gamma} \epsilon^{\alpha' \beta' } \partial_{\beta '} \omega_{\alpha `} \;, \;\;
    R_{\beta \delta} = \frac{1}{\sqrt{g}} g_{\beta \delta} \epsilon^{\alpha \gamma} \partial_{\gamma} \omega_{\alpha} \;, \;\;
    R = \frac{2}{\sqrt{g}} \epsilon^{\alpha \beta } \partial_{\beta} \omega_{\alpha} \;,
\end{equation}
from which it follows
\begin{equation}\label{a12}
    {R^{\alpha \beta}}_{\gamma \delta} =  \frac{1}{2} \epsilon^{\alpha \beta} \epsilon^{\delta \gamma} R \;, \;\;
    R_{\beta \delta} = \frac{1}{2} g_{\beta \delta} R \;.
\end{equation}
From these the scaling properties of $\omega_\alpha$, $R$ and $R_{\alpha \beta}$ are
\begin{equation}\label{alor}
{\omega'}_\alpha = \omega_\alpha - \sqrt{g} \epsilon_{\alpha \beta} \sigma^\beta \;, \;\;
    e^{-\sigma} {R'}_{\alpha \beta} - R_{\alpha \beta} = g_{\alpha \beta} \Box_g \sigma \;, \;\;
    e^{\sigma} {R'} - R = 2 \Box_g \sigma \;,
\end{equation}
where primed objects refer to the metric ${g'}_{\alpha \beta} = e^{2 \sigma} g_{\alpha \beta}$ and
$\Box_g \sigma = \frac{1}{\sqrt{g}} \partial_\alpha (\sqrt{g} \partial^\alpha \sigma)$. Notice that the scaling properties of the Ricci and scalar curvature are consistent with $R_{\alpha \beta} = \frac{1}{2} g_{\alpha \beta} R$ (multiply both sides of the equation for the Ricci by $e^{-\sigma} g^{\alpha \beta}$). These scaling properties differ slightly from those reported in \cite{Iorio:1996ad} (see,e.g., Eq. (28) there). While for the applications of those results to the case in point this discrepancy is relevant, hence we took it into account in our calculations here, it is not so for the conclusions of \cite{Iorio:1996ad} because there the important considerations refer to the flat space limit for which, e.g., the last relation in (\ref{alor}) becomes $\Box \sigma = 0$, which is the same constraint obtained in \cite{Iorio:1996ad} and the correct condition for conformal symmetry in $n=2$.

Notice that for $g_{\alpha \beta} = - \delta_{\alpha \beta} e^{-2 \sigma}$ it is $\omega_\alpha = e^{-2 \sigma} \epsilon_{\alpha \beta} \sigma^\beta$, hence from the last relation in (\ref{a11})
\begin{equation}\label{a13}
     R = 2 e^{2 \sigma} \triangle \sigma \;.
\end{equation}

Explicit form of the action in the conformally flat case:
\begin{equation}\label{aexplicitaction}
i \int d^n x \sqrt{g} \bar{\Psi} \gamma^a E_a^\mu \nabla_\mu \Psi = i \int d^n x \sqrt{g} \; \bar{\Psi} \gamma^a E^\mu_a (\partial_\mu + \frac{1}{2} \omega_\mu^{\; b c} J_{b c}) \Psi \;,
\end{equation}
when
\begin{equation}\label{metricandfield}
    g_{\mu \nu} = e^{-2 \sigma (x)} \eta_{\mu \nu}  \;,
\end{equation}
then
\begin{equation}\label{variaconfflat}
    e^a_\mu = e^{-\sigma} \delta^a_\mu \;, E_a^\mu = e^{\sigma} \delta_a^\mu \;, \sqrt{g} = e^{- n \sigma} \;,
\end{equation}
hence from (\ref{a7})
\begin{equation}\label{aomegacf}
    {\omega_\mu}_{b c} = \delta^a_\mu (\eta_{a b} \delta^\nu_c - \eta_{a c} \delta^\nu_b) \sigma_\nu \;.
\end{equation}
From these and from
\begin{equation}\label{alorentz}
    \gamma^a J_{a b} = \frac{n - 1}{2} \gamma_b \;,
\end{equation}
it is easy to see that the action is
\begin{equation}\label{afinal}
    i \int d^n x \; e^{- (n -1) \sigma} \; \bar{\Psi} \gamma^a (\partial_a - \frac{n - 1}{2} \sigma_a) \Psi \;.
\end{equation}
When we use $\Psi = e^{\frac{n-1}{2} \sigma(x)} \Psi'$ the action reduces to the flat one $i \int d^n x \bar{\Psi}' \not\!\partial \Psi'$.

Let us prove here that $S^\sigma (x_1, x_2) = e^{\frac{n-1}{2} (\sigma (x_1)+ \sigma (x_2))} S ' (x_1, x_2)$ satisfies
\begin{equation}\label{apppropeq}
    i \not\!\nabla_{x_1} S^\sigma  (x_1, x_2) = \frac{1}{\sqrt{g}} \delta^{n} (x_1 - x_2) \;.
\end{equation}
First apply the flat Dirac operator to $S^\sigma$
\begin{equation}
    i \not\!\partial_{x_1} S^\sigma  (x_1, x_2) = i \frac{n - 1}{2} \gamma^a \sigma_a (x_1) S^\sigma  (x_1, x_2)
+ e^{(n -1) \sigma (x_1)} \delta^{n} (x_1 - x_2) \;,
\end{equation}
or
\begin{equation} \label{aprefinal}
    i \gamma^a (\partial_a -  \frac{n - 1}{2} \sigma_a)(x_1) S^\sigma  (x_1, x_2) = e^{(n -1) \sigma (x_1)} \delta^{n} (x_1 - x_2) \;,
\end{equation}
where we used $i \not\!\partial_{x_1} S ' (x_1, x_2) = \delta^{n} (x_1 - x_2)$. Then use the metric in (\ref{metricandfield}) and (\ref{variaconfflat})-(\ref{alorentz}) to write (\ref{aprefinal}) as
\begin{eqnarray}
i \gamma^a e^{\sigma(x_1)} (\partial_a -  \frac{n - 1}{2} \sigma_a)(x_1) S^\sigma  (x_1, x_2) & = &
i \gamma^a E_a^\mu (\partial_\mu -  \frac{n - 1}{2} \sigma_\mu)(x_1) S^\sigma  (x_1, x_2) \\
= e^{n \sigma (x_1)} \delta^{n} (x_1 - x_2) & = & \frac{1}{\sqrt{g}} \delta^{n} (x_1 - x_2) \;,
\end{eqnarray}
i.e. equation (\ref{apppropeq}).

\end{document}